\documentstyle[aps,multicol,epsf,epsfig,aps,rotate]{revtex}

\begin{document}

\draft

\title{Discrepancy between Monte-Carlo Results and Analytic
  Values \\ for the Average Excluded Volume of Rectangular Prisms}  

\author{Sameet Sreenivasan,$^\ast$ Don R. Baker,$^{\ast\dagger}$ Gerald
  Paul,$^\ast$ and H. Eugene Stanley$^\ast$}

\address{$^\ast$Center for Polymer Studies and Dept. of Physics,
Boston University, Boston, MA 02215 USA\\ 
$\dagger$Department of Earth and Planetary Sciences,
McGill University\\ 3450 rue University, Montr\'eal, QC H3A 2A7 Canada}

\date{29 July 2002}

\maketitle

\begin{abstract}

We perform Monte Carlo simulations to determine the average excluded
volume $\langle V_{\mbox{\scriptsize ex}}\rangle$ of randomly oriented
rectangular prisms, randomly oriented ellipsoids and randomly oriented
capped cylinders in 3-D. There is agreement between the
analytically obtained $\langle V_{\mbox{\scriptsize ex}}\rangle$ and the
results of simulations for randomly oriented ellipsoids and randomly
oriented capped cylinders. However, we find that the $\langle
V_{\mbox{\scriptsize ex}}\rangle$ for randomly oriented prisms obtained
from the simulations differs from the analytically obtained results. In
particular, for cubes, the percentage difference is $3.92$,
far exceeding the bounds of statistical error in our simulation. {\bf Added in
Revision 2: We recently found the cause of the discrepancy between the
simulation result and the analytic value of the excluded volume to be the
effect of an error in our simulation code. Upon rectification of the
simulation code, the simulation yields $ 11.00 \pm 0.002 $ as the excluded
volume of a pair of randomly oriented cubes of unit volume. The simulation
also yields results as predicted by the analytic formula for all other cases
of rectangular prisms that we study.}

\end{abstract}

\bigskip

\begin{multicols}{2}

The concept of excluded volume is used widely in the statistical
mechanics of gases as well as polymer systems. The excluded volume of an
object is defined as the volume around an object into which the center
of another similar object is not allowed to enter if overlapping of the
two objects is to be avoided\cite{Onsager49}. In the case of objects
that are allowed random orientations in a specified angular interval
one defines an average excluded volume $\langle V_{\mbox{\scriptsize
ex}}\rangle$ that is the excluded volume averaged over all possible
orientational configurations of the two objects. 

The excluded volume arises in the leading order concentration expansion
(``virial expansion'') for the pressure in the case of gases that repel
each other with a hard-core volume exclusion\cite{Garboczi}. In polymer
systems, the concept arises due to the fact that in dilute polymer
solutions, each polymer molecule tends to exclude all others from the
volume that it occupies \cite{Flory53}.  In studying continuum
percolation with a system of soft-core objects, one of the properties of
interest is the connectivity of the percolating cluster. The mean number
of intersections per object is a convenient measure of the connectivity
and is given by the number of object centers located within the average
excluded volume of an object. Hence, the product of the critical
concentration $N_c$ of objects at the percolation threshold and the
average excluded volume $\langle V_{\mbox{\scriptsize ex}}\rangle$ gives
the critical average number of intersections per object $B_c$
\cite{Balbergetal84,Balberg87,Alonetal90} 
\begin{equation} B_c = N_c
\langle V_{\mbox{\scriptsize ex}}\rangle.  \label{equation1}
\end{equation} 
The behavior of $B_c$ has been the focus of previous literature
\cite{Balbergetal84,Balberg87,Alonetal90} and we recently found this
quantity to be approximately invariant for systems of a given object
shape independent of orientational constraints in 2-D\cite{SS02}. In this
paper we discuss our simulation results for the $\langle
V_{\mbox{\scriptsize ex}}\rangle$ of rectangular prisms, ellipsoids of
revolution and capped cylinders. We find a statistically significant
difference between our simulation results and the analytically
predicted values for the $\langle V_{\mbox{\scriptsize ex}}\rangle$ of
rectangular prisms while obtaining an agreement, within statistical
error, for the case of ellipsoids and capped cylinders.

In order to find the average excluded volume for a given object shape we
employ a Monte Carlo simulation algorithm similar to the one used in
Ref. \cite{Garboczi}.  An object of the shape under consideration is
placed with its center coinciding with the center of a box of side $L$ . The
box volume is chosen to be larger than the excluded volume, but small
enough to minimize the number of wasted trials. The orientation of the
object is specified by the three Euler angles: $0 \le \phi \le 2\pi$, $0
\le \theta \le \pi$, $0 \le \psi \le 2\pi$.  The object is given an
initial orientation that is random.  In order to achieve random isotropic orientations, the first
and last Euler angle are drawn from a uniform random distribution while
the second Euler angle is drawn from a cosine distribution. A second
identical object is then introduced into the box with its center
randomly positioned in the box and given a random orientation. We
then determine if the two objects intersect. We repeat this procedure
for $N_T=10^{9}$ trials and record the number of times $N_I$ the two
objects intersect. The probability that the two objects intersect,
$P_I$, is $N_I/N_T$. The average excluded volume for a pair of objects
oriented randomly is $P_IL^3$, where $L^3$ is the volume of the box. We carry
out the simulation for rectangular prisms and ellipsoids of different
aspect ratios, as well as for capped cylinders. The method we use to
determine the intersection of rectangular prisms is described in detail
in Ref.~\cite{Bakeretal02}. We determine the intersection of ellipsoids
using a contact function \cite{JVBaron}, and the algorithm to check the
intersection of two capped cylinders is in Ref.~\cite{bourke89}

The analytic expression for the average excluded volume for a pair of convex
bodies labeled A and B is \cite{Kihara53}
\begin{equation}
\langle V_{\mbox{\scriptsize ex}}\rangle = (V_{A} + V_{B} +
(M_{A}S_{B})/4\pi + (M_{B}S_{A})/4\pi),
\label{equation2}
\end{equation}
where $V$, $S$ and $M$ denote the volume, surface area and the total
mean curvature of the two bodies respectively.

For two identical convex bodies $\langle V_{\mbox{\scriptsize
ex}}\rangle = 2(V + (MS)/4\pi)$. Reference~\cite{Kihara53} contains a
derivation of Eq.~(\ref{equation2}) as well as the expressions for V, M
and S for different shapes. For a rectangular prism with sides $l_{1}$,
$l_{2}$, $l_{3}$,
\begin{equation}
  M = \pi(l_{1}+l_{2}+l_{3}).
\end{equation}
The analytic expression for $\langle V_{\mbox{\scriptsize ex}}\rangle$
can also be obtained using a well known result in geometric probability
known as the ``principal kinematic formula''
\cite{Santalo,Schneider,Klain}. The formula gives the measure of the set
of positions of the center of one object for which it intersects another
fixed object which, apart from a normalization factor, is
Eq.~(\ref{equation2}).

Table I contains our simulation results for $\langle V_{\mbox{\scriptsize
ex}}\rangle$ for a pair of identical rectangular prisms compared with
the analytically obtained $\langle V_{\mbox{\scriptsize ex}}\rangle$
for different aspect ratios of the prisms. The simulation results do not agree
with the expected analytic values, e.g., the 
simulation result exceeds the analytic result for cubes by 3.92\%. We see that
the percentage difference decreases as the length of one side becomes
much greater than the other two (i.e., as the prism approaches a
widthless stick), while the percentage difference increases when the
length of one side becomes smaller than the other two (i.e., as the
prism tends to a platelet). Simulations for soft core rectangular prisms
have also been performed in Ref.\cite{Saar02}. The value
of $\langle V_{\mbox{\scriptsize ex}}\rangle$ for cubes quoted in
Ref.~\cite{Saar02} is $ 10.56$, in good agreement with our simulation
result although the authors of Ref.~\cite{Saar02} were unaware of the
discrepancy with the analytically obtained value of $11$ .

We reproduced the results earlier derived by Garboczi {\it et. al.}     
\cite{Garboczi} for the average excluded volume of two identical
ellipsoids of revolution, for different aspect ratios. Table II shows
our simulation results for $\langle V_{\mbox{\scriptsize ex}}\rangle$ of
ellipsoids of revolution of different aspect ratios compared with the
analytically obtained $\langle V_{\mbox{\scriptsize ex}}\rangle$
\cite{Garboczi}. As found in Ref.~\cite{Garboczi}, we see that the
simulation results for both prolate and oblate ellipsoids agree with the
analytic results, to within statistical error. This is in
contrast to the case of rectangular prisms.

Our simulation results for the $\langle V_{\mbox{\scriptsize
ex}}\rangle$ of capped cylinders compare well, within 
statistical error, with the analytically-obtained values for these objects
\cite{Balbergetal84}. The capped cylinders in our simulation consist of
a cylinder of length $L = 1.0$ whose ends have hemispherical caps of
radius $R = 0.25$. Our simulation results yield $\langle
V_{\mbox{\scriptsize ex}}\rangle=2.8797 \pm 0.0006$ while the
analytically determined value is $2.8798$. The procedure in
Ref.\cite{Balbergetal84} to determine $\langle V_{\mbox{\scriptsize
ex}}\rangle$ of two capped cylinders consists of finding the excluded
volume for a given relative orientation between the two, and then
calculating the average value of the expression over all possible
orientations of both objects. We verify that the analytic result in
Ref.~\cite{Balbergetal84} can also be obtained using Eq.~(\ref{equation2}).

Finally we performed a simulation to determine the $<V_{ex}>$ of a sphere and a cube. Our simulation result 
agrees with the value obtained analytically: for a cube and a
sphere, both of unit volume, the excluded volume using
Eq.(\ref{equation2}) is 9.349 while the simulation yields $9.347
\pm 0.001$. The analytical result for this case can be
determined using the method employed in Ref.\cite{Balbergetal84} and we find
exactly the same result as the one obtained using Eq.~(\ref{equation2}).
 
We have not determined the cause of the difference between the
simulation results and the analytic results for $<V_{ex}>$ of rectangular
prisms. We have, however, carried out checks to try to eliminate the
the possibility of certain sources of error in our simulations. Possible
sources of error in the simulation are :

\begin{itemize}

\item[{(i)}] The simulation does not generate random isotropic orientations
  of the objects.
\item[{(ii)}] The algorithm used to determine the intersection of the
  rectangular prisms fails to do so accurately. 
\item[{(iii)}] The number of trials in the simulation is insufficient to
obtain good statistics.
\end{itemize}
Possibility (i) appears to be ruled out by the agreement, within
statistical error, of the results of our simulation for ellipsoids and
capped cylinders and their analytically-predicted values. As a check on
the intersection algorithm, we carry out a simulation to determine the $\langle
V_{\mbox{\scriptsize ex}}\rangle$ for parallel aligned cubes which yields
a value of $8.000 \pm 0.002$ , in good agreement with the expected value of 8
\cite{Balbergetal84}. We also determine the excluded volume for a number
of fixed relative orientations of the prisms and confirm agreement
with the corresponding analytic results. Therefore possibility (ii) also
appears to be eliminated. As far as possibility (iii) is concerned, we
find that there is no significant change in the value $<V_{ex}>$ after
about $10^{7}$ trials and we, therefore, believe that $10^{9}$ trials
yield good statistics. There are instances where Monte Carlo
simulations yield results different from those obtained by exact
enumeration, such as in the study of self-avoiding walks in percolation
\cite{Anke}, where failure to sample certain configurations has an
inordinately strong effect on the final result. However, we cannot
identify any such configurations in our study of $<V_{ex}>$.

 There may also be other sources of error, which we have not
considered. However, with regard to this, we point out that both our
simulation result for the $<V_{ex}>$ of cubes as well as the simulation
result in  Ref.~\cite{Saar02} are in excellent agreement and were
obtained independently and without knowledge of each other's work at the time.

Looking at our results, we observe that there is agreement between the
simulation and analytic results for the $<V_{ex}>$ of ellipsoids which
are non-singular objects (smooth) and have only a single point of contact when touching
externally. We find agreement in the case of capped cylinders which
are non-singular objects, but which can have multiple points of contact when
touching externally. Also, in the case of the sphere and cube we obtain good
agreement between simulation and analytic results. In this case only one
of the objects viz. the cube is singular. In the case of two rectangular
prisms for which we find the discrepancy, both the objects are
singular. Hence, an initial conjecture as to the cause of the difference
could be the possibility that Eq.(\ref{equation2}) does not hold for a
pair of convex bodies, both of which have singular surface points. 
Eq.(\ref{equation2}) is derived in Ref.\cite{Kihara53} beginning with the
assumption that the objects under consideration are non-singular and
have a single point of contact when they touch externally. Thus, the
derivation should not apply for objects such as rectangular
prisms. After Eq.~(\ref{equation2}) is obtained, the assumption of
smoothness is dropped with minimal explanation and Eq.(\ref{equation2})
is applied to singular objects such as rectangular prisms \cite{Kihara53}.  
 
On the other hand, the principal kinematic formula is widely
accepted to be true and the literature on the subject
\cite{Santalo,Schneider,Klain} states that the expression must work for
{\it all\/} convex bodies. Furthermore, derivations of the principal
kinematic formula seem not to make any assumptions about the
smoothness of the bodies under consideration. 

The difference of our Monte-Carlo results from the analytically
predicted values is intriguing and because of the broad applications of
the concept of excluded volume, the problem merits further study.

\bigskip
\centerline{\bf EPILOGUE} 

 {\bf We recently found the cause of the discrepancy between the
simulation result and the analytic value of the excluded volume to be the
effect of an error in our simulation code. Upon rectification of the
simulation code, the simulation yields $ 11.00 \pm 0.002 $ as the excluded
volume of a pair of randomly oriented cubes of unit volume. The simulation
also yields results as predicted by the analytic formula for all other cases of
rectangular prisms that we study.}

\subsubsection*{Acknowledgments} 

We thank Intervep, NSERC and NSF for support. We also thank Professor
Paul Goodey at the University of Oklahoma for providing useful
clarifications and encouragement.

\end{multicols}

\newpage

\begin{table}
\label{Table1}
\caption{Comparison of simulation results with analytic results for the
average excluded volume of rectangular prisms of unit volume. The sides
of the prisms are $l_{1}$,$l_{2}$ and $l_{3}$ with $l_{2} = l_{3}$. The
aspect ratio of the prism is defined as $l_{1}/l_{2}$.
We estimate the uncertainty in $\langle V_{\mbox{\scriptsize ex}}\rangle$
as follows: The reciprocal of the square root of the number of Monte
Carlo trials yielding intersection of the two objects is the fractional
uncertainty in the determination of $\langle V_{\mbox{\scriptsize
ex}}\rangle$. The product of the fractional uncertainty and the
estimated value of $\langle V_{\mbox{\scriptsize ex}}\rangle$ is the
uncertainty in that value.}
\begin{tabular}{cclcc}
\hline
Aspect Ratio & Monte Carlo & Calculated &
Absolute Difference & Percentage Difference \\
\hline
Platelets \\
0.000001 & $1751600.0 \pm 467.911$  & 2000002.000 & 248402.0 & 12.420 \\
0.01 & $182.760 \pm 0.048$ & 207.020 & 24.260 & 11.718\\
0.5 & $11.488 \pm 0.002$ & 12.000 & 0.513 & 4.272\\
0.6 & $11.058 \pm 0.002$ & 11.533 & 0.475 & 4.118\\
0.7 & $ 10.804 \pm 0.001$ & 11.257 & 0.453 & 4.027\\
0.8 & $10.659 \pm 0.001$ & 11.100 & 0.441 & 3.975\\
0.9 & $10.591 \pm 0.001$ & 11.022 & 0.431 & 3.911\\  
\hline
Cube \\
1 & $10.569 \pm 0.001$ & 11.0 & 0.431 & 3.918\\
\hline
Prisms \\
2 & $11.543 \pm 0.003$ & 12.0 & 0.477 & 3.811\\
4 & $15.023 \pm 0.003$ & 15.500 & 0.477 & 3.077\\  
8 & $22.774 \pm 0.007$ & 23.250 & 0.476 & 2.047\\
16 & $38.654 \pm 0.016$ & 39.1250 & 0.471 & 1.204\\
32 & $70.624 \pm 0.046$ & 39.12500 & 0.439 & 0.617\\
64 & $134.599 \pm 0.175$ & 135.03125 & 0.432 & 0.320\\
\end{tabular}
\end{table}    
\newpage
\begin{table}
\label{Table2}
\caption{Comparison of simulation results with analytic results for the
average excluded volume of randomly oriented unit volume oblate and
prolate ellipsoids of revolution. The lengths of the axes of the
ellispoids are a,b and c, with $b = c$. The aspect ratio of the ellipsoid
is defined as a/b. The uncertainty in $\langle V_{\mbox{\scriptsize
ex}}\rangle$ is estimated as in Table I.}
\begin{tabular}{lcccc}
\hline
Aspect Ratio & $\langle V_{\mbox{\scriptsize ex}}\rangle$ (Monte Carlo) &
$\langle V_{\mbox{\scriptsize ex}}\rangle$ (Analytic result) \\ 
\hline
Oblate \\ 
0.03125 & $77.739  \pm 0.006$ & 77.741 \\ 
0.0625 & $40.281 \pm 0.003$ & 40.281 \\
0.125 & $21.811 \pm 0.002$ & 21.810 \\
0.25 & $12.956 \pm 0.001$ & 12.956 \\ 
0.5 & $9.077 \pm 0.007$ & 9.077 \\
\hline
Sphere \\
1 & $8.001 \pm 0.002$ & 8.0 \\                   
\hline
Prolate \\
2 & $9.077 \pm 0.001$  & 9.077 \\
4 & $12.957 \pm 0.002$ & 12.956 \\
8 & $21.813 \pm 0.005$ & 21.810 \\
16 & $40.28 \pm 0.014$ & 40.281 \\
32 & $77.70  \pm 0.04$ & 77.741 \\
\end{tabular}
\end{table}
\end{document}